\documentclass{article}

\usepackage{arxiv}

\usepackage[utf8]{inputenc} 
\usepackage[T1]{fontenc}    
\usepackage[bookmarksnumbered=true]{hyperref} 
\hypersetup{
     colorlinks = true,
     linkcolor = blue,
     anchorcolor = blue,
     citecolor = blue,
     filecolor = blue,
     urlcolor = blue
     }
\usepackage{url}            
\usepackage{booktabs}       
\usepackage{amsfonts}       
\usepackage{nicefrac}       
\usepackage{microtype}      
\usepackage{lipsum}
\usepackage{lscape}
\usepackage{graphicx}
\usepackage{lineno}
\usepackage{array}
\usepackage{longtable}
\usepackage{amsmath}

\title{Trends in the optimal location and sizing of electrical units in smart grids using meta-heuristic algorithms}

\author{
  Kayode E. Adetunji \\
  School of Electrical and Information Engineering\\
  University of the Witwatersrand\\
  Johannesburg\\
  \texttt{kayvins@gmail.com} \\
   \And
  Ivan Hofsajer \\
  School of Electrical and Information Engineering\\
  University of the Witwatersrand\\
  Johannesburg\\
  \texttt{ivan.hofsajer@wits.ac.za} \\
     \And
  Ling Cheng \\
  School of Electrical and Information Engineering\\
  University of the Witwatersrand\\
  Johannesburg\\
  \texttt{ling.cheng@wits.ac.za} \\
}

\begin{document}
\maketitle

\begin{abstract}
The development of smart grids has effectively transformed the traditional grid system. This promises numerous advantages for economic values and autonomous control of energy sources. In smart grids development, there are various objectives such as voltage stability, minimized power loss, minimized economic cost and voltage profile improvement. Thus, researchers have investigated several approaches based on meta-heuristic optimization algorithms for the optimal location and sizing of electrical units in a distribution system. Meta-heuristic algorithms have been applied to solve different problems in power systems and they have been successfully used in distribution systems. This paper presents a comprehensive review on existing methods for the optimal location and sizing of electrical units in distribution networks while considering the improvement of major objective functions. Techniques such as voltage stability index, power loss index, and loss sensitivity factors have been implemented alongside the meta-heuristic optimization algorithms to reduce the search space of solutions for objective functions. However, these techniques can cause loss of optimality. Another perceived problem is the inappropriate handling of multiple objectives, which can also affect the optimality of results. Hence, a recent method such as Pareto fronts generation has been developed to produce non-dominating solutions. This review shows a need for more research on (i) the effective handling of multiple objective functions, (ii) more efficient meta-heuristic optimization algorithms and/or (iii) better supporting techniques.
\end{abstract}

\section{Introduction}\label{section1}
The field of optimization has developed quickly in the past few years. Optimization problems occur in many fields such as physics, biology, engineering, economics, commerce, management science, and even politics. In engineering, areas are spread around process control, characterization, approximation theory, curve fitting, modelling, which are in the field of civil, electrical, chemical, and mechanical. In this cause, several optimization solutions have evolved to tackle related optimization problems. Some general methods are numerical, experimental, analytical, and graphical methods. Numerical methods have been the most prominent of these methods, and regarded as mathematical programming \cite{Peralta2012,Borwein2013a}. Over time, mathematical programming (basically deterministic or exact) has spanned into several branches such as linear, non-linear, quadratic, dynamic, integer, mixed integer linear, and mixed integer non-linear programming. These methods have been termed as deterministic or exact methods.

Current developments in optimization algorithms have seen a major division into two types which are deterministic and heuristic. Deterministic algorithms are developed such that they find an optimal solution at a logically required time \cite{Yang2011a}. They use the features of the problem to generate a search graph where convergence to optimum values is attained. However, the tougher problems may pose higher computational requirement such as time and memory. Most problems fall in the category of NP-hard or NP-complete, as they increase exponentially in dimensions to a discrete change in variables. Heuristic algorithms on the other hand, are designed to greedily search a possible best solution to such puzzling problem, usually having superiority (in terms of computational efficiency) over the deterministic algorithms \cite{Chun1998a}. These heuristic algorithms are problem-dependent and are often stuck at a local optima. This means that the solution found may be worse compared to other potentially available viable solutions. This effect may also cause computational complexity. Overcoming this shortcoming birthed the meta-heuristic algorithms.

Meta-heuristic algorit+hms are dynamic algorithms that extensively search for a solution in global optima \cite{Yang2011a}. They are independent of any problem and their high-level nature permits them to roam in and out several local optima. Meta-heuristics are mostly nature-inspired and are majorly based on two concepts: evolutionary and swarm-based intelligence algorithms \cite{Venter2010a}. Evolutionary algorithms are started with the initialization of a random population, where the best individuals are passed to the next generation (clearly imitating the theory of evolution). Swarm-based intelligent algorithms mimic the social behavior of animals and how they collectively interact with each to achieve a common goal. Hence, meta-heuristic algorithms have been applied heavily to the field of power systems, and to be more specific, distribution networks of smart grids.

In order to get full benefits of the integration of units in smart grids, their location and size should be optimally determined on the basis of better power/voltage quality, reduced power loss, and improved investment cost. This has led to several reviews. \cite{Wong2019} and \cite{Das2018} reviewed different technologies and benefits of Energy Storage Systems (ESS), and methods for optimal location, sizing and control. The authors emphasized on further studies to be done on the performance and control of ESS. \cite{Askarzadeh2017} reviewed the application of a meta-heuristic algorithm, harmony search in power systems. Their studies were focused on economic dispatch/unit commitment, optimal power flow, control, optimal placement of FACTS devices, expansion and planning, prediction, parameter identification, reconfiguration, optimal reactive power dispatch among others. \cite{Sirjani2017b} reviewed different methods used in the optimal placement and sizing of Distribution Synchronous Static Compensator (D-STATCOM). It was concluded that there is need to place D-STATCOM and in different conditions, and improve on the speed and accuracy in solving the optimal placement and sizing of D-STATCOM. \cite{Sheibani2018} reviewed ESS installation and expansion in distribution and transmission networks. A qualitative analysis was carried out on the storage type, objective functions, constraints, and solutions. \cite{Latreche2018} and \cite{Prakash2016} focused on different approaches used for solving the optimal location and sizing of DG units, considering the objective functions, indices, and constraints. \cite{Sujil2018} reviewed multi agent system applications to power system problem. They focused on its application to DG units management system, electric vehicle management system, electricity market, energy management and control, power generation expansion, and fault detection and protection.

Till date (2019), meta-heuristic algorithms have been heavily applied to the optimal location and sizing of units in a distribution system. Therefore, there is a need to review existing research studies on the application of these algorithms in smart grids. This paper focuses on the pros and cons in the application of meta-heuristic algorithms to smart grids planning, especially with the optimal sizing and location of electrical units. The rest of this paper is as follows: \textcolor{blue}{Section \ref{section2}} discusses the brief background of smart grids main constituents and meta-heuristic algorithms. \textcolor{blue}{Section \ref{section3}} presents a detailed review of  meta-heuristic algorithm application in optimal sizing and location of energy sources and devices for distribution systems in smart microgrids. \textcolor{blue}{Section \ref{section4}} discusses the summary of findings from reviews. Finally, conclusions are drawn in \textcolor{blue}{Section \ref{section5}}.

\section{Background}\label{section2}
\subsection{Smart grids}
Microgrids are replica of a grid system in a constrained situation, which can be case of a remote location or a standalone institution. However, the development of a microgrid must have the standard essential components such as DG, storage facility, and load. These components are connected through the Point of Common Coupling (PCC), and can be controlled by switches for dynamic energy source use, or connection to a main grid. Hence, a microgrid can be islanded or standalone, and can be an AC or DC network. The characteristics of a microgrid can also permit only low and medium voltage distribution network system \cite{Hatziargyriou2005a}.

Smart grids have a bi-directional communication system which allows for the dynamic control and communication for updates in the distribution network. The National Institute for Standards and Technology (NIST) have coined out concepts from a standard smart grid, which are architecture, architecture process, energy services interface, functional requirement, harmonization, interchangeability, and inter-operability \cite{NIST2012}. One significant device in the topology of the smart grid is smart meters \cite{Cardenas2012a}, which could be central for control, sensing, and communication \cite{Aggarwal2010a}. The installation makes it seamless for data communication.  The input of Renewable Energy Systems (RES) in smart grids comes with a limitation, as its intermittent nature will not allow continuous power flow in the distribution network. Therefore, Energy Storage systems (ESS) have been deployed to overcome this shortcoming. ESS makes it possible to store energy for high demand hours, which promotes the demand side management in a smart grid. Another property of an ESS is its allowance for a smoothing power output through frequency variation \cite{Li2017} and \cite{Indu2016a}. 

\begin{figure}
\centering
\includegraphics[scale=0.5]{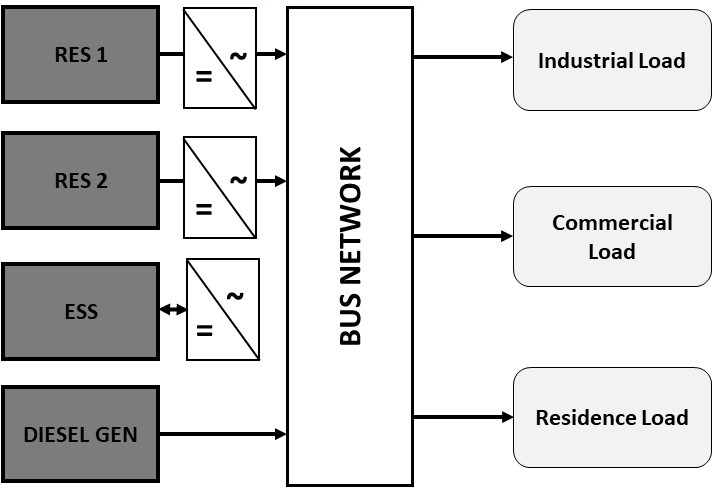}
\caption{Configuration of a microgrid with the integration of an ESS (adapted from \cite{Patrao2015a}) \label{fig1}}
\end{figure}

An integration of ESS in a microgrid is shown in \autoref{fig1} \cite{Patrao2015a, Mohd2015, Palizban2016a}. However, the implementation of this technology comes with a cost \cite{Oudalov2007a,Wade2010a,Makarov2012a, Zakeri2015a}. Sizing and siting of ESS are the major economic factors of developing a smart grid. Optimal sizing eliminates initial cost and redundant energy \cite{Aghamohammadi2014a}, while optimal siting minimizes power line loss, which may in turn cause an increase in size of the ESS or even the whole generators. Therefore, the location of an ESS cannot be overemphasized. \cite{Chiradeja2005a} emphasized on the importance of the line loss reduction, and included the location and rating of generators as important factors to line loss reduction.

There are major parameters that are considered when finding an optimal size and location in a distribution network. These are the real power of a bus, the reactive power from a bus, the resistance on a branch due to real power flow, and the reactance on a branch due to reactive power flow. The power flow/load flow studies make use of these parameters to mostly calculate the total power loss in a network. More discussion is in \textcolor{blue}{Section \ref{2.2.1}}. Given the potential of standalone smart grids, this paper presents a review on the recent meta-heuristic algorithms used for optimal sizing and location of key units that constitute a smart grid.
\subsubsection{Power Flow Model }\label{2.2.1}
\begin{figure}
\centering
\includegraphics[scale=0.60]{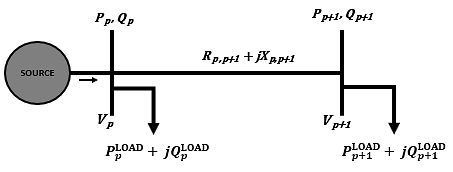}
\caption{Schematic of a single line illustration of an RDN \label{fig2}}
\end{figure}

Power flow is an important factor in distribution systems. In order to measure the transmission or distribution loss in a network, an appropriate representation of power flow must be achieved. Power flow methods range from optimal power flow (OPF), continuous power flow (CPF), probabilistic power flow (PPF) and so on. These methods have been used to analyze the components on a power bus line, hence determining the calculation of power losses on such lines.  However, power loss equations will be based on the type of generator source. For example, the output power of diesel generators will defer from the power from PVs or WT. Also, single- or three-phase type will change the model of its power flow. The derivation for the single-phase power supply is shown from the initial diagram in \autoref{fig2} \cite{MohamedImran2014a}. Since current is a flow of electricity on a power line, it plays a huge role in the calculation of power flow.

\begin{figure}[b]
\centering
\includegraphics[scale=0.60]{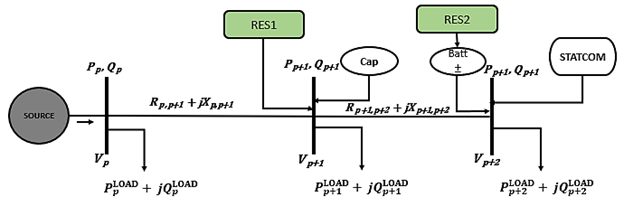}
\caption{Addition of devices to a single line illustration of an RDN \label{fig3}}
\end{figure}

%

\begin{equation}\label{eqn1}
P_{p,p+1} = P_p - P_{p+1}^{load} - {R_{p,p+1}\frac{(P_{p,p+1}^2 - Q_{p,p+1}^2)}{|V_P|^2}}
\end{equation}
\begin{equation}\label{eqn2}
Q_{p,p+1} = Q_p - Q_{p+1}^{load} - X_{p,p+1}\frac{(P_{p,p+1}^2 - Q_{p,p+1}^2)}{|V_P|^2}
\end{equation}
\begin{equation}\label{eqn3}
\begin{split}
    |V_{p,p+1}|^2 = |V_p|^2 - 2(R_{p,p+1}^2P_{p,p+1}^2+X_{p,p+1}^2Q_{p,p+1}^2) + (R_{p,p+1}^2+X_{p,p+1}^2)
\frac{(P_{p,p+1}^2 + Q_{p,p+1}^2)}{|V_P|^2} 
\end{split}
\end{equation}
where$ p$ is the sending bus and $p+1$ is the receiving bus. The total power loss of the system is represented in \autoref{eqn4} below

\begin{equation}\label{eqn4}
P^{LOSS}_{total} =  \displaystyle\sum_{p=1}^{n-1} P^{loss}_{p,p+1}
\end{equation}
According to the \autoref{eqn4}, derived loss equations can be supplemented by the addition of energy devices such as capacitors and STATCOM. \autoref{fig3} shows the addition of battery and a wind turbine, which are factors of voltage instability. Capacitors are used for supporting power flow through reactive power enhancement. On the other hand, STATCOM devices consists of Voltage Source Converters (VSCs) coupled with transformers and energy devices. They are used for compensating bus voltage distribution power systems, hence improving power quality. STATCOM dynamically injects and/or absorbs reactive power for improving voltage stability and profile \cite{Sirjani2017b}.

With the addition of other energy sources. \autoref{eqn5} - \ref{eqn7} may be updated as follows:
\begin{equation}\label{eqn5}
P_{p+1} = P_p - P^{LOAD}_{p+1} - ({R_{p,p+1}\frac{(P_{p,p+1}^2 - Q_{p,p+1}^2)}{|V_P|^2}})+P^{RES1}+P^{BATT} 
\end{equation}
\begin{equation}\label{eqn6}
Q_{p+1} = Q_p - Q_{p+1}^{LOAD} - (X_{p,p+1}\frac{(P_{p,p+1}^2 - Q_{p,p+1}^2)}{|V_P|^2})+\alpha_qQ^C_{p+1} 
\end{equation}
The reactive power can be updated as shown in \autoref{eqn7}
\begin{equation}\label{eqn7}
Q_{p+2} = Q_{p+1} + Q^{STATCOM}_{p+2} 
\end{equation}
$P_{RES1}$ and $P_{RES2}$ can be integrated as solar modules and battery energy systems respectively. Here, $\alpha Q^C$ is the reactive power compensation by the capacitor with an $\alpha$ factor and $Q^{STATCOM}_{p+2}$ is the reactive power compensation by the STATCOM at bus ${p+2}$.

The aforementioned units need a proper sizing and location to avoid excess installation and maintenance cost and cascading effect on a power system network respectively. The cascading effect might largely affect the sizing of any units, thereby hitting on the economic value of a smart grid. These effects explain the importance of optimization in a power system network.

\subsection{Meta-heuristic Algorithms}
The concept of most meta-heuristic algorithms is based on agent or set of agents (called multi agent systems). These systems are comprised of characterized agents that operate without human interaction, to achieve an objective or a set of objectives \cite{Pipattanasomporn2009}. An agent has the ability to coordinate an action based on the current situation of the system and/or other agents within that system. Particle Swarm Optimization (PSO) \cite{Kennedy1995} and Ant Colony Optimization (ACO) \cite{Dorigo2005} are good examples of the use of agents. On the other hand, there are population based meta-heuristic algorithms such as the Genetic Algorithm (GA) \cite{Davis1991}, which uses the Darwinian evolution and the natural selection theory. The defined characteristics of a meta-heuristic algorithm is well related to its underlying inspiration, which simulates the behavioural routine without being monitored. This concept makes it intelligent, dubbed computational intelligence (a part of artificial intelligence).

\begin{flushleft}
\begin{figure}
	
	\includegraphics[scale=0.50]{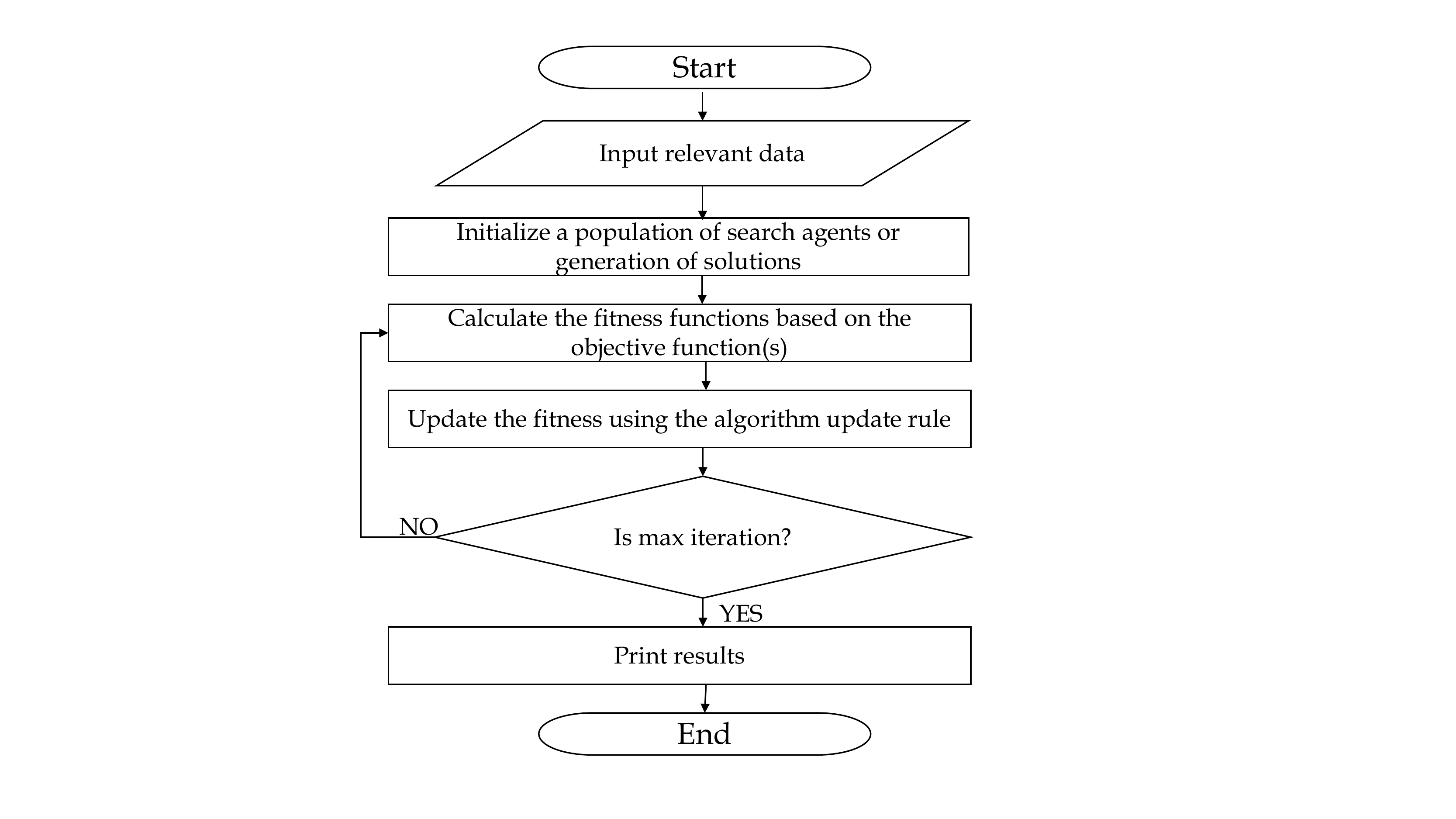}
	\caption{Generic flowchart of a meta-heuristic algorithm \label{fig4} \cite{Wong2019}}

\end{figure}
\end{flushleft}

A standard approach for searching for best solutions are based diversification and intensification \cite{Yang2010}. The former is a sporadic search of a whole solution search, while the latter is the search of a particular region of the search space. This concept is the fundamental for reaching a global optima, hence finding optimal solutions. Procedural measures to solve the objective functions are illustrated in \autoref{fig4}. 

The superiority of a particular algorithm to solve every problem is unrealistic. This is substantiated by the ``No Free Lunch'' theory \cite{Wolpert1997}, which states that all algorithms will perform averagely. This means that algorithm A can outperform algorithm B in a specific objective function, but algorithm B may outperform algorithm A in a different objective function. This has made researchers focus on the problem, rather than finding the overall best algorithm for every problem. The CEC benchmark functions incorporate single and multi-objectives and constrained objectives \cite{Chen2005}. These functions are used to test newly developed or modified algorithms.

The advent of the benchmark functions have seen the development of other algorithms such as Bat Algorithm (BA) \cite{Yang2010d}, Cuckoo Search (CS) \cite{Yang2010c}, Harmony Search (HA) \cite{Loganathan2001}, Flower Pollination Algorithm (FPA) \cite{Yang2010b}, and Firefly Algorithm (FA) \cite{Yang2010a}. Other new meta-heuristic algorithms have also been developed such as Ant Lion Optimizer (ALO) \cite{Mirjalili2015}, Whale Optimization Algorithm (WOA) \cite{Mirjalili2016}, Grey Wolf Optimizer (GWO) \cite{Mirjalili2014}.

One major benefit of meta-heuristic algorithms is their capability to solve highly computational tasks at a substantial efficiency. These algorithms will present a near best result rather than using highly computational resources for a perfect result. In applications such as power systems, the problem can be combinatorial which may make it an NP-hard, making deterministic or analytical approaches not advisable. Also, continuous nature of power system such as sizing needs to be very ``near best'' to avoid extreme economic badness. Therefore, local optima  (where results may be far away from being the optimum best) is not entertained. Since decisions are stochastic, there is a likelihood that such algorithms hit the global optima during a rerun. However, A good reason for meta-heuristic inappropriate setting of parameters can be a premature convergence \cite{RezaeeJordehi2013a}.

The application of both new and old meta-heuristic algorithms in power systems, specifically optimal placement and sizing of electrical units has seen a massive success.

\section{Optimal Location and Sizing of Electrical Units}\label{section3}

\subsection{Energy Sources}
Energy sources are central to the development of smart grids. These energy sources can be Renewable Energy Sources (RES) and may be configured as a single source or multiple DG units. Either type of energy source configuration should have an appropriate size and location of the energy source itself. Some examples of energy sources used in literature are Battery Energy Storage Systems (BESS), wind turbines, PV modules, and diesel generators. BESS are used to compensate for the epileptic nature of the RES. While BESS have been enhanced with improvement in chemical composition and structure \cite{Das2018}, there is also a need to optimally place them at strategic location to achieve optimum power delivery. 

Researches such as in  \cite{KhalidMehmood2017,Ramirez2018a,Rodriguez-Gallegos2018a,Dong2016a} proposed algorithms to find optimal size and location of BESS in a distribution networks. Individual energy sources such as PVs and wind turbines have also been optimally placed and sized for optimum energy transfer (\cite{Dixit2016a}). However, these sources can be DG units and their optimal placement in a distribution network is essential for power delivery. The authors \cite{Das2018a,Jani2018a,Dehghanian2013a,Moradi2016a,Boktor2018a} developed meta-heuristic algorithms to optimally size and place BESS-based DG units in distribution networks.

\subsection{Power Electronic Devices}
The inception of microgrid comes with the power delivery problem, which is mostly power stability. Power electronic devices have been used to compensate distorted power, thereby enhancing power quality. Overtime, power electronic devices have been sized and placed at strategic location. Examples are capacitors (or capacitor banks), Distribution Synchronous Static Compensator (D-STATCOM), and Dynamic Voltage Restorer (DVR). Capacitors are generally the most economic, hence it attracts more studies. Researches such as in \cite{MohamedShuaib2015a, Abdelaziz2016b, Montazeri2019, Ali2016a, Prakash2017b, Upper2018a, Mendoza2019a} worked on the optimal sizing and allocation of capacitors in a distributed network through the use of meta-heuristic algorithms. Other power electronic devices such as D-STATCOM, which consists of coupling transformers, energy storage devices and inverters \cite{Modarresi2016}, have also been optimally placed and sized.

\subsection{Non-arbitrary Data-set}
Optimal sizing and placement of energy sources and power electronic devices are based on parameters such as real \& reactive power of buses and resistance \& reactance of branches that connect the buses. These parameters influence the major objective functions such as power loss, voltage profile, and voltage stability. Bus and branch (or line) datasets are used for evaluating algorithm performance. The IEEE dataset is a standard for solving sizing and placement problems, and has been strongly prioritized for choosing research papers for this review. Real data from location grids were also considered from literature. 

\subsection{Metaheuristic Algorithms for Objective Optimization}

Algorithms such as the GA and PSO are more like a paradigm for meta-heuristic algorithms, as they have been successfully applied in different applications. This explains the prevalence of their applications in optimal sizing and location. Some of their applications are discussed below.

\cite{Shwehdi2018a} utilized the GA to acquire the optimal data for load flow analysis. The method was used to enhance reactive power flow to optimal placement of capacitor banks in strategic locations along bus lines, and allowed for different load scenarios, hence its combinatorial nature. The developed algorithm was tested on the Saudi Electricity Company (SEC). In the same vein, \cite{Liu2019} considered the optimal placing of BESS in a Virtual Power Plant (VPP) for optimal planning of a distribution network. The objective was to reduce the power loss and the power fluctuation induced by PV plants, while considering uncertainties of power output and load growth. The GA was synchronized with a Monte Carlo simulation to achieve optimal planning. However, the algorithm is not computationally efficient, mainly because of the slow convergence time. 

\cite{DeAraujo2018a} used a strongly modified GA to enhance power loss minimization objective, for optimal capacitor placement in an unbalanced distribution system. They considered daily load variation curve as their loading conditions, and tested the algorithm on the IEEE 4- and 123-bus and 85-bus feeder. Multiphase Optimal Power Flow (MOPF) tool was used to find near-best allocation, while considering reactive power injections. GA was utilized for optimum compensation values of the reactive power on the bus, with the use of discrete capacitors. 

In the quest for constraint handling, \cite{Vuletic2016a} developed a genetic algorithm that specially handles constraint for discrete optimization. This kind of optimization technique would be efficient for solving the optimal placement of capacitors. The algorithm could handle constraints without using a penalty function, which eliminates the process of selecting a penalty parameter, hence it was coined as a Penalty Free Genetic Algorithm (PFGA). Three sets of load profile (residential, commercial, and industrial) were loaded as active and reactive power. The PFGA was used to minimize power and energy losses, and was tested on 18-, 68-, and 141-bus systems.

From \cite{Xiao2015}, the GA was implemented alongside the modelling of the OPF for the optimal siting and sizing of BESS in a distribution network. The OPF method specifically enhances BESS scheduling and minimizes network losses while the GA optimizes the net present cost (NPV). The algorithm was tested on the IEEE 33-bus system, and was tested with previous GA implementation. The results show that the number of generations of a GA can be reduced while achieving good results, hence reducing computational time.  

Recently, \cite{Singh2019} developed a GA-based optimization that was implemented for the integration of DGs, STATCOM, and Plug-in Hybrid Electric Vehicles (PHEVs). This was done with optimal location and size of the energy sources and a FACTS device, to reduce total real power loss in the distribution network. The DGs and loads were modeled into four types to savor the real-world characteristics. The GA evaluates the fitness values of each chromosome, which in turn updates each bus power state. The algorithm was evaluated on the IEEE 37-bus RDN, with conclusion that the integration of DGs, STATCOM, and PHEVs will deliver enhanced real and reactive power support and enhanced system power factor. The framework can be evaluated on other test bus systems. Other meta-heuristic algorithms may also be implemented for performance comparison. PSO has been implemented in the same light.
 
\cite{Mosbah2017a} used the PSO for optimally locating and sizing of shunt capacitors in a radial distribution system to reduce high current that causes voltage drop and to minimize real line power loss. The algorithm was tested on an IEEE 10-, 15-, and 34-bus RDN.

\cite{Lee2015a} proposed a discrete PSO for the optimal placement of capacitors, considering load patterns in distribution systems. A Gaussian probabilistic distribution, with a chaotic model was implemented for power flow. This model enhances the voltage profile and minimizes the power loss in a distribution network. For further studies, algorithm should be evaluated on bus test systems. In the same vein, \cite{Karimi2016a} used the PSO for solving comprehensive objective function to optimally place and size capacitor banks. Real and reactive power were considered in the reduction of power loss, and a function was used to penalize line voltage drops. The algorithm was tested on a 10-, 33-, and 69-bus distribution system. 

\cite{Das2018a} presented a method to optimally place distributed BESSs in a distribution network using Artificial Bee Colony (ABC), with improvements being made through addressing power loss alongside voltage deviation and line loading. A heavily renewable energy-dependent IEEE 33-bus system was used as a testbed for the simulations and also for comparing the ABC to the PSO.  From \cite{Injeti2015a}, two metaheuristic algorithms (Bat Algorithm (BA) and Cuckoo Search (CS)) were compared, while solving the optimal capacitor sizing and location problem. The objective was to minimize the real power loss and maximize network savings, which was evaluated on a 34- and 85 bus system. It was concluded that the CS is better than the BA in solution quality, but slower to converge than BA.

\cite{Saha2016a} formulated a multi-objective problem based on power loss minimization, voltage deviation maximization, and voltage stability improvement. A Chaos Symbiotic Organisms Search (CSOS) algorithm was developed to optimally place and size DGs in a microgrid based on the objectives, and was tested on 33-, 69-, and 118-bus RDS. The algorithm was also compared with the conventional Symbiotic Organisms Search (SOS). 

Recently, a lot of algorithms have been developed to improve the objective functions regarding optimal sizing and location. Some are the Whale Optimization Algorithm (WOA), Ant Lion Optimizer (ALO), Moth Flame Algorithm (MFA) and Grey Wolf Optimizer (GWO). Some of their successful applications in optimal placement and sizing are itemized below:

\begin{enumerate}

	\item \cite{George2018a} used ALO-based optimization technique to optimally place fixed shunt capacitors in a distribution system, with the consideration of objective functions such as minimization of both total distribution power loss and total annual cost. A backward/forward technique was used to compute the load flow of the test system, and was tested on the IEEE 33- and 69-bus test system. \cite{Boktor2018a} also utilized the ALO for optimal placing of shunt capacitors in an RDN, with objectives to minimize power loss and to improve voltage profile. The ALO was evaluated on IEEE 33-bus and IEEE 69-bus.


	\item \cite{Prakash2017b}  implemented WOA for optimal sizing and placement of capacitors in an RDN. Their objective was to reduce power loss, to improve voltage profile, and to minimize cost. The WOA was compared to other meta-heuristics such as PSO and BFOA, and was tested on IEEE 34-bus and IEEE 85-bus system.  

	\item \cite{Sirjani2017c} used the Discrete Lightening Search Algorithm (DLSA) for optimal placement of capacitors in wind farms. The proposed objectives were based on energy loss minimization and management cost reduction. However, power loss must be calculated to solve for energy loss. The DLSA was compared to the GA and Discrete Harmony Search Algorithm (DHSA). 

	\item \cite{Sanjay2017a} developed a hybrid Grey Wolf Optimizer for optimally allocating DGs in a microgrid. Their objective was to reduce power loss (both active and reactive). They evaluated the algorithm on the IEEE 33-bus, IEEE 69-bus, and the Indian 85-bus radial system. 
	
	\item From \cite{Ceylan2017a}, MFA was used to optimally size and place capacitor banks in distribution networks. Power flow was solved for loss minimization using an iterative algorithm while an arbitrarily simulated real and reactive load profile with a 15-minute interval was used for the load conditioning. 
	
	\item \cite{Moazzami2017a} used Modified Shuffled Frog Leaping Algorithm (MSFLA) for optimally locating and sizing of DGs and D-STATCOM. Their objective was to minimize distribution line losses and increase voltage stability, to improve power quality. They performed the algorithm on the IEEE 33-bus system, and compared it to the GA algorithm.

\end{enumerate}
So far, the efficient use of meta-heuristic is dependent on the mode of handling objective functions. Multi-objective functions stand the risk of not being optimized simultaneously due to the distinctive interference among them. For example, a form of multi-objective handling (sequential goal programming) will create a master and slave objective, where the master objective is prioritized. During iterations, the master objective function is handled firstly, which paves a way to handle the slave objective(s). This process is repeated until all objectives are handled \cite{AbouELEla2009}. It is observed that this method of multi-objective handling can be computationally expensive among other drawbacks. Another type of multi-objective handling is aggregation of more objectives into a single objective. This is done through weights assignment. The aggregated objective is solved a priori to the optimization process. However, this method may require (i) multiple runs with weights variation to achieve optimal solutions, (ii) a good knowledge of objective prioritization (which is always subjective), and (iii) scaling for different objectives. In addition, there is also no information exchange among solutions during the optimization process and Pareto fronts are not feasible due to the non-convex outputs. Most of these drawbacks can be obviated by the a posteriori method of multi-objective handling.

The a posteriori method is a Pareto-based multi-objective function handling that eliminates the setback of conflicting objectives by generating an optimum set of points called the Pareto frontier. All objective functions are optimized collectively and the non-conflicting solutions are selected as the best outputs \cite{Pindoriya2010}. These outputs can be non-dominating, which implies that a subset of solutions is not entirely better than other subset of solutions. Fewer studies regarding the a posteriori method have been carried out in the field of optimal location and sizing of electrical units. Some of the studies are discussed below:

\cite{Dehghanian2013a} proposed a special multi-objective, non-dominated sorting genetic algorithm (NSGA-II) for the optimal siting of DG units in a power system, with objective functions to reduce network power losses and to increase system reliability. Probabilistic method was used to determine the power flow and uncertainties with the consideration of constraints and uncertainties. The algorithm was tested on an IEEE 37-bus system. \cite{Jannat2016a} used similar algorithm for the optimal capacitor placement in distribution networks. The authors implemented an RE-based power flow with the consideration of uncertainties from active power of wind and solar energy and load. However, the objective was voltage profile improvement and was evaluated on the Serbia grid network.

\cite{KhalidMehmood2017} used the NSGA-II for computing optimal solutions for battery energy storage system (BESS) sizing and allocation. They used the objective function to estimate aggregate the energy losses in the distribution network, and the total investment cost of DGs and BESSs. By using voltage regulation, they increased the lifespan of the BESS. The algorithm was tested on an IEEE 906 bus European test feeder. \cite{Farsadi2016a} used NSGA-II for power loss minimization and voltage profile improvement. Electricity prices and probabilistic load (with peak) were modelled based on time series for optimally sizing and placing capacitors in a distribution system. The algorithm was tested on the IEEE 33-bus distribution test system. 

\cite{Mahesh2016a} also implemented the PSO in a non-dominated sorting multi-objective as an advanced Pareto front. The algorithm was to minimize total power loss and improve voltage profiles, while sizing and placing DGs optimally in an RDN. The algorithm was implemented alongside the VSI and PLR techniques.

	\begin{longtable}{>{\centering\arraybackslash}p{1.7cm}>{\centering\arraybackslash}p{1.2cm}>{\centering\arraybackslash}p{2cm}p{1.5cm}p{1.4cm}p{2cm}p{2cm}}
		\caption{Optimization with no indices\label{table1}}\\
		\hline\hline
		\multicolumn{7}{c}%
		{}\\
		\textbf{References} & \textbf{Grid scenario} & \textbf{MH algorithm} & \textbf{Flow model type} & \textbf{unit type} & \textbf{Objective} & \textbf{Evaluation system} \\
		\hline 
		\endfirsthead
		\caption[]{Optimization with no indices (continued)}\\
		\hline\hline
		\multicolumn{7}{l}%
		{}\\
		\textbf{References} & \textbf{Grid scenario} & \textbf{MH algorithm} & \textbf{Flow model type} & \textbf{Unit type} & \textbf{Objective} & \textbf{Test system} \\
		\hline
		\endhead
		\hline  \\
		\cite{Shwehdi2018a} & Transmission inter-tie & GA & OLF & Capacitors & Reactive power flow enhancement & 37-bus system \\ 
		
		\cite{DeAraujo2018a} & Unbalanced DS & GA & MOPF & Capacitors & Power loss minimization & IEEE 4- and 123-bus system \\ 
		
		\cite{Vuletic2016a} & DS & Penalty-free GA & HPF & Capacitors & Power loss minimization & 18- and 69-, 141-bus system \\ 
		
		\cite{Xiao2015} & RE-based DS & GA & OPF & DG and ESS & network loss minimization, enhanced ESS scheduling & IEEE 33-bus system \\ 
		
		\cite{Liu2019} & VPP & GA & PPF & BESS & ESS cost reduction, power deviation improvement, cost reduction & IEEE 33 test feeder \\ 
		
		\cite{Singh2019} & smart grid & GA & OPF & DG, STATCOM, and PHEV & Total real power loss minimization & IEEE 37-bus system \\ 
		
		\cite{KhalidMehmood2017} & RE-based DS & NSGA-II & PPF & WT, PV, and BESS & Energy loss minimization, Voltage profile improvement, cost minimization & IEEE 906 bus European low-voltage test feeder \\ 
		
		\cite{Farsadi2016a} & DS & NSGA-II & PPF & Capacitors & Power loss minimization, Voltage profile improvement & IEEE 33-bus system \\ 
		
		\cite{Dehghanian2013a} & DS & NSGA-II & PPF & DG & Network loss minimization, Cost minimization & IEEE 37-bus system \\ 
		
		\cite{Jannat2016a} & RE-based DS & NSGA-II & OPF & Capacitors & Voltage profile improvement & Serbia real RDN \\ 
		
		\cite{Mosbah2017a} & DS & PSO & OPF & Capacitors & Power loss minimization, total annual cost minimization, & IEEE 10- and 15-, 34-bus system, Algerian real RDN \\ 
		
		\cite{Lee2015a} & DS & Discrete PSO & Chaotic LF & Capacitors & Power loss minimization, Voltage profile improvement & IEEE 33-bus system \\ 
		
		\cite{Karimi2016a} & DS & PSO & OPF & Capacitors & Power loss minimization & IEEE 10-, 33-, and 69-bus system \\ 
		
		\cite{Das2018a} & RE-based DS & ABC & OPF & ESS & Power loss minimization, Voltage profile improvement & IEEE 33-bus system \\ 
		
		\cite{Injeti2015a} & DS & BA \& CS & B/F sweep PF & Capacitors & Real power loss minimization and network savings maximization & 34-bus and 85-bus system \\ 
		
		\cite{George2018a} & Microgrid & ALO & B/F sweep PF & Capacitors & Power loss minimization and total annual cost minimization & IEEE 33-bus and 69-bus test systems \\ 
		
		\cite{Boktor2018a} & DS & ALO & B/F sweep PF & Capacitors & Power loss minimization and Voltage profile improvement & IEEE 33-bus and 69-bus test systems \\
		
		\cite{Vuletic2014a} & DS & ALO & B/F sweep PF & Capacitors & Power loss minimization and Voltage profile improvement & IEEE 33-bus and 69-bus test systems \\
		
		\cite{Sirjani2017c} & Wind Farms & Discrete LSA & OPF & Capacitors & Energy loss minimization and Voltage profile improvement & 40-bus test systems \\
		
		\cite{Sanjay2017a} & DS & FPA & OPF & DG & Power loss minimization and Voltage profile improvement & IEEE 33-bus, 69-bus, and Indian 85-bus test systems \\
		
		\cite{Ceylan2017a} & DS & MFA & OPF & Capacitors & Power loss minimization and Voltage profile improvement & 33-node feeder \\
		
		\cite{Moazzami2017a} & DS & MSFLA & B/F sweep PF & DG and D-STATCOM & Power loss minimization and Voltage profile improvement & IEEE 33-bus test systems \\
		
		\cite{Upadhyay2018a} & DS & PSO-GA & N-R PF & Capacitors & Power loss minimization and Voltage profile improvement & 34-bus test systems \\
		
		\cite{Sedghi2016} & DS & TS-PSO & PPF & ESS & Power loss minimization and Voltage profile improvement & 21-node test feeder \\
		
		\cite{Prakash2017b} & DS & WOA & OPF & Capacitors & Power loss minimization and Voltage profile improvement & IEEE 34-and 85-bus test systems \\
		\hline
	\end{longtable} 
\subsubsection{Optimization with indices method}
Voltage stability is another concern in a power systems network. There is a need for a power system to maintain satisfactory level of voltage at the buses with regards to normal or disturbing conditions. Such conditions can be varying loads or power injections from distributed sources. As per the IEEE/CIGRE ({Canada2009}) definition, ``voltage stability refers to the ability of a power system to maintain steady voltages at all buses in the system after being subjected to a disturbance from a given initial operating condition''. An extreme case of voltage instability is a voltage collapse. This may be due to bad weather events or overloaded power lines. 

The inability to compensate for reactive power loss will also cause voltage instability (\cite{Modarresi2016}), especially with small-scale network microgrids. This is because the load type and varying cause a negative effect on voltage stability. Reactive power, $ Q $ moves from the low voltage zone to the high voltage zone. Hence, there would be a need for high voltage to transmit reactive power over long distances. However, the upsurge of reactive power during transmission will increase loss of active and reactive power ($P_{Loss}$ and $Q_{Loss}$).

Voltage Stability Index (VSI) is a method used for controlling voltage instability. This can be carried out in two ways: either by identifying the set of weakest buses and lines in a distribution network or by adding a reverse component in real time \cite{Modarresi2016}. In the first case, a static analysis can be done by obtaining power system data to evaluate for the weak buses, then the set of weakest buses is identified for the placement of DGs, capacitors, D-STATCOM, or BESSs. The second case is achieved through a wide area measurement system (WAMS), which consists of phasor measurement units that provide necessary data in real time for fending off voltage instability. 

Since VSIs can be used alone to solve planning of distribution systems, several methods have been developed with deviating strategies (\cite{Modarresi2016}). Some of the proposed methods are Line Stability Index (Lmn), Line Stability Index (Lp), Novel Line Stability Index (NLSI), Line Voltage Stability Index (LVSI), Fast Voltage Stability Index (FVSI), Voltage Collapse Proximity Index (VCPI) ect. Overtime, VSI has been combined with meta-heuristic algorithms for distribution network planning. A common equation for solving VSI is given in \autoref{eqn8}:
\begin{equation}\label{eqn8}
\text{VSI}_{(p+1)}= V_p^4-4[P_{(p+1)} X_p- Q_{(p+1)} R_p ]^2-4[P_{(p+1)} R_p+ Q_{(p+1)} X_p ]^2 V_p^2 
\end{equation} 
Another index is the Power Loss Index (PLI), which has been developed to serve as a measure for power loss on a transmission or distribution line. The PLI will be calculated through the measure of power loss reduction at every node, and buses with larger PLI will have the priority to be selected as candidate buses for possible compensation. Studies that have used these indices are discussed in the following.

From \cite{AbulWafa2014a}, a fuzzy GA was implemented alongside VSI to particularly enhance voltage stability for optimal placement and sizing of capacitors in a DS. The algorithm was tested on a 33-node RDN.

\cite{Moradi2016a} utilized the combination of GA and Intelligent Water Drops (IWD) for optimally sizing and allocating DGs in a microgrid. VSI was used to reduce active power losses through the identification of candidate buses. The hybrid algorithm was tested on a 33-bus and 69-bus system and shows a good computational time which increases linearly with number of DGs. 

From \cite{Poornazaryan2016a}, a new index based on the VSI method, was successfully implemented with a modified Imperialistic Competitive Algorithm (ICA) for optimally placing and sizing DG units. The algorithm minimizes real and reactive power loss and improves the voltage profile in different load scenarios, and it was tested on a 34-bus and 69-bus test system. Comparison with the CS algorithm showed improvement in the voltage profile and reactive power loss.

\cite{El-Ela2018a} used a newly developed algorithm, Water Cycle Algorithm (WCA) to simultaneously size and place DGs and capacitors in a microgrid. Their algorithm was proposed as single- and multi-objectives, to minimize distribution power loss and to improve voltage profile through the VSI. Economic and environmental factors such as reduction of generation costs and emission reductions were also considered. 

From \cite{Jani2018a}, a hybrid multi-objective algorithm (Multi-Objective Particle Swarm Optimization (MOPSO) and NSGA-II), was proposed to optimally allocate energy storage systems in a wind farm grid, considering the uncertainties. Their objective was to improve voltage deviation and to reduce operation costs and carbon emissions. The proposed algorithm was tested on the IEEE 30-bus system.

The NSGA-II algorithm was implemented in \cite{Celli2018} to optimally allocate and size BESS in a distribution system. A probabilistic load flow method was used to model a 24-hour steady and emergency state configuration with the objective function to minimize real power losses. The NSGA-II was tested on an unknown rural distribution network. \cite{Khaki2018} proposed a new framework for simultaneously placing and sizing of wind turbines (WT) and BESS in a distribution network. The framework is based on the GA using a probabilistic approach to simulate wind power and BESS output. Their objective was to minimize total system loss and the cost of WTs, which was tested on the IEEE 33-bus system. 

\cite{Zhang2018} used Chance Constrained Programming (CCP) to solve the probabilistic power flow in the optimal planning of distribution systems. The planning involves the siting of DGs at the commencement of distribution networks. The objective was to reduce economical cost through the correlation of uncertainties using NSGA-II for the Pareto optimal fronts. Wind speed, illumination intensity, and load profiles were considered for the uncertainty parameters, and a 61-bus test system was used for evaluation. However, no comparison was made with other algorithms to test for performance. It is also observed that the method will not be applicable with the use of BESS in distribution systems.

\cite{Mostafa2019} implemented the Symbiotic Organism Search (SOS) for improving the performance of microgrids through the optimal allocation of ESSs. The SOS was implemented alongside the VSI method to identify the most sensitive nodes to critical voltage instability. The whole algorithm was based on daily curve and renewable DGs power output to minimize power loss, to improve voltage profiles and to boost voltage stability of microgrids. The algorithm was not evaluated on a test bus system neither was it benchmarked with other algorithms.

\cite{Ali2016a} proposed the Improved Harmony Algorithm (IHA) for optimally sizing capacitors in an RDN. The PLI was used to determine possible buses for the optimal installation of capacitors, and was followed by the implementation of the IHA. The algorithm was tested on a 69-bus distribution system, and was compared to other algorithms such as the PSO, ABC, DE, and HS. 

From \cite{El-Fergany2013a}, the PLI was used to detect high potential buses for effective injection of reactive power. Afterwards, the CS algorithm was used to optimally place shunt capacitors in distribution networks, to reduce peak power loss and to improve voltage profiles. The performance of the CS algorithm was examined on a 33-bus and 69-bus system. 

From \cite{Abdelaziz2016b}, Flower Pollination Algorithm (FPA) and PLI was implemented to solve the optimal location and size of capacitors in an RDS. The algorithm was tested on 15-bus, 69-bus, and 118-bus RDS, and was compared with the PSO, DSA, TBLO, ABC, CS, HSA, and Plant Growth Simulation Algorithm (PGSA). 

\cite{Reddy2017a} used the WOA to optimally size RE-based DG units for the minimization of power losses, improvement of voltage profile, and increase of reliability in an RDN. An index vector was used to select candidate buses for optimal location of the DG units, and was verified with other types of DG units with varying power factor. The whole algorithm was evaluated on the IEEE 15-, 33-, 69-, and 85-bus system. 

\cite{Montazeri2019} proposed a new PLI method for identifying possible buses placing capacitors. They implemented the crow search algorithm to solve the combinatorial problem of capacitor placement. The algorithm was tested on a 69- and 118-bus test system and was compared to other variants of the PLI technique.

\cite{Lotfi2019} proposed a hybrid algorithm to optimally allocate capacitors for the reconfiguration of distribution feeders fed with ESS, DG and solar PV. An improved PSO and MSFLA was used alongside with VSI technique to achieve power loss minimization and voltage deviation reduction. The VSI technique was used as one of the objective functions, following a load flow technique from Thevenin’s equivalent circuit. The VSI is used with a penalty factor for unstable decision parameters, hence avoiding buses that have VSI values greater than zero. The IPSO-MSFLA algorithm was evaluated on the IEEE-95-nodes test system. However, the modified algorithm was not compared with other algorithm.

\cite{RoyGhatak2018} introduced an extended version of the NSGA-II (E\_NSGA-II) to optimally add solar PV, BESS, and D-STATCOM to a smart microgrid, using a probabilistic model for power flow. A VPI technique was used to compute voltage profile improvement after the integration of the three units. The algorithm was tested on a 69-bus test system to evaluate the voltage profile, environmental benefit, reliability, and benefit cost ratio. Afterwards, a non-parametric test was performed to compare the performance of the proposed algorithm to other multi-objective algorithms such as MOGA, MOPSO, and NSGA-II.

\subsubsection{Optimization with Loss Sensitivity Factors}
Loss sensitivity factors (LSF) is a method to determine candidate nodes for the optimal allocation and sizes of DGs and capacitors. These factors are attained from the parameters derived from power flow models. LSF can be used to reduce both real and reactive power, especially in the case of DGs, where reactive power can also be supplied. The calculation for LSF for both active and reactive power loss is given in \autoref{eqn9} and \autoref{eqn10}.
\begin{equation}\label{eqn9}
\frac{\partial Ploss_{(m,m+1)}}{\partial P_{(m+1)}}  = R_{(m,m+1)}\frac{2Q_{(m+1)}}{V_{(m+1)}^2}    
\end{equation}
\begin{equation}\label{eqn10}
\frac {\partial Qloss_{(m,m+1)}}{\partial Q_{(m+1)}} = X_{(m,m+1)}\frac{2Q_{(m+1)}}{V_{(m+1)}^2 }    
\end{equation}
Studies from literature pertaining to the use of LSF with meta-heuristic algorithms show that real power loss is mostly considered. A loss sensitivity matrix is attained as in \autoref{eqn11}.
\begin{equation}\label{eqn11}
\begin{pmatrix}\frac{\partial Ploss}{\partial P_2} & \frac{\partial Qloss}{\partial P_2} \\
\frac{\partial Ploss}{\partial Q_2} & \frac{\partial Qloss}{\partial Q_2}
\end{pmatrix}
\end{equation}

\begin{figure}
\centering
\includegraphics[scale=0.4]{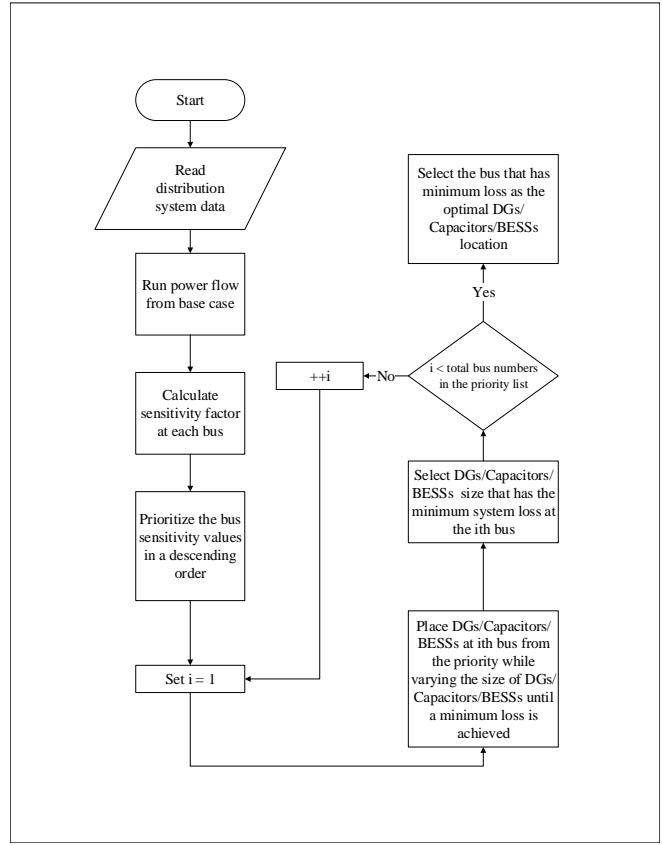}
\caption{Flowchart of an LSF algorithm as a standalone optimizer \label{fig6}}
\end{figure}

 The LSF technique also helps to reduce the search space of candidate solutions and the values are stored in a priority list (in a descending order). These values can be based on normalized voltages of 1.01 or lesser, and are calculated by dividing the voltage on each bus by 0.95. A flowchart of LSF algorithm is illustrated in \autoref{fig6}.
 
\cite{Elsheikh2014a} implemented the PSO alongside the LSF to solve the optimal capacitor and sizing problem in an electric power system. Their framework used the LSF technique to select candidate buses that require compensation on a power line. Afterwards, the PSO was implemented as a discrete form to determine the optimal size of capacitors. The proposed framework was evaluated on a 10-, 15-, and 34-bus system, and results show improvement in voltage profile and line loss minimization. 

\cite{Diab2017a} developed a hybrid PSO (coupled with Quasi Newton algorithm) for the optimal location and size of shunt capacitors in distribution systems. The objective function was to enhance voltage profile, to minimize power loss, and to decrease total annual system cost. The LSF technique was implemented to obtain the most suitable buses for capacitor placement, and was followed by the implementation of the hybrid PSO to choose the optimal location. The algorithm was examined on the IEEE 15-bus and 34-bus standard RDS, and the 111-bus Moscow grid. 

\cite{Mendoza2019a} optimally sized and allocated capacitors in a distribution network using PSO, to reduce real power losses and economic costs. The Backward/Forward sweep power flow method was used to derive the relevant parameters. The developed PSO was evaluated on 34-bus and 85-bus system was mainly compared with the WOA. Results show that the PSO had higher voltage profile improvement than the WOA.

\cite{Sharma2019a} proposed a hybrid ABC-PSO approach with load flow calculation based on fuzzy load flow to solve capacitor sizing, to minimize power loss and to improve voltage profile. The LSF technique was used to detect buses sensitive to power loss, then a fuzzy inference system was used to select the optimal capacitors placement. ABC-PSO algorithm was used to size the capacitors was tested on a 34-node RDN. Also, \cite{Muthukumar2018a} developed a hybrid algorithm, HSA-ABC to optimize capacitor size and placement in an RDN, considering different load models. PLI, VSI, and LSF were implemented, to calculate total network power loss, to detect low quality voltage in nodes, and to identify high active power loss in nodes for capacitor placement respectively.

\cite{MohamedShuaib2015a} worked on the optimal location and size of capacitors in a distribution network. Here, the LSF technique was implemented to select candidate buses for the capacitor placement, which led to the implementation of the Gravitational Search Algorithm (GSA) to size optimally size the capacitors on the selected buses. From \cite{Abdelaziz2016c}, LSF was used to assign candidate buses with lowest values for capacitor placement. The FPA-based algorithm was used to optimally select the LSF-elected buses for capacitor placement, with the focal objective function to minimize real power loss. The algorithm was evaluated on a 10-, 33-, and 69-bus system. 

\cite{Upper2018a} developed a MFA-based algorithm to optimally size capacitors in RDNs, with the objective to reduce energy losses considering variation of load while using the LSF technique to select the possible candidate buses. The results from the LSF technique preceded the implementation of the MFA. The algorithm was tested on a standard 69-bus RDS. From \cite{Ali2018a}, ALO was implemented alongside LSF to optimally allocate and size renewable DGs in a microgrid respectively. The proposed algorithm was evaluated on a 69-bus RDS and compared to other algorithms to show the improvement of total power loss reduction and net savings enhancement. 

\cite{Kishore2018a} developed an improved bacterial foraging optimization algorithm (IBFOA) with symmetric fuzzy methods to optimally place and size capacitors in RDN. LSF and VSI were implemented alongside the algorithm to minimize power loss and improve voltage stability. Their framework was tested on a 33-, 69-, and 141-node RDN. 

\cite{Li2018} used a two-stage optimization framework to optimally place and size BESS and DGs in an active distribution network. The framework consists of an LSF technique and a multi-objective ALO (MOALO), which solved the initial capacity and location of the DGs and BESS respectively. The MOALO was initially used to find Pareto-optimal solution, which is followed by obtaining the order of significance of each Pareto solutions. The final results addresses the objective which minimizes the power losses and maximizes the voltage stability and investment benefits, while considering the uncertain outputs of energy sources (DG and BESS). The framework was tested on the PG \& E 69-bus and compared to the NSGA-II, MOPSO, and MOHA. Results showed that their two-stage optimization method is better than the aforementioned algorithms in terms of line losses voltage stability and investment costs. 

Similarly, \cite{Babacan2017} optimally sized and placed BESS in a PV-integrated grid distribution system, using a GA-based bi-level optimization framework. The aim was to reduce voltage fluctuations caused by PV outputs to the RDN. Voltage fluctuations could be termed as voltage instability since its effect can also break down a PSN. The study was validated on the IEEE 8500-node test feeder and was compared to an evolutionary algorithm. Their work may also be compared to other meta-heuristic algorithms, to test for fast convergence, accuracy, and computational time. Also, this method may also be carried out to minimize line losses.

\section{Discussion}\label{section4}
The summary of studies on distribution systems is presented in Tables \ref{table1}, \protect\ref{table2}, and \ref{table3}. The studies focused on the optimal sizing and placement of energy sources and power electronic devices. The tables were categorized according to the techniques (such as Voltage Sensitivity Index (VSI), Power Loss Index (PLI), and Loss Sensitivity Factor (LSF)) supplemented to meta-heuristic algorithms. Each tables were categorized into objective functions, meta-heuristic algorithm types, grid scenarios, and distribution test systems types.  

The study uncovers studies on simultaneous sizing and placement of two or more capacitors, STATCOM, DGs, and BESS. DGs have been classified into types such as (i) real power supply, e.g. from PVs (ii)  reactive power supply, e.g. from capacitors and STATCOMs (iii) real power supply and absorbs reactive power e.g. wind turbines (iv) real and reactive power supply e.g. combined heat and power (CHP) plants. However, the size and location of these DGs will affect voltage stability and investment cost. Since all DGs do not supply reactive power, power electronic devices compensate for steady voltage profile, hence finding optimal solutions. 

In the era of promoting clean energy and making up for the intermittent power supply from RE, optimal sizing and allocation of ESS in active RDNs have been understudied. It is evident from the review that probabilistic and fuzzy load flow model has been developed in lieu of earlier models such as DLF, OPF and backward/forward power flow.

From the review, major applied meta-heuristic algorithms are the PSO and GA. These algorithms have been modified and combined with other meta-heuristic algorithms for increased efficiency such as in \cite{Upadhyay2018a,Sedghi2016,Sharma2019a,Lotfi2019}. However, new algorithms have also been applied and bench-marked for performance evaluation. Such studies are from \cite{Prakash2017b,Ali2018a,Reddy2017a,El-Ela2018a,Montazeri2019}. Traditional PSO and GA have been improved and outperformed the new algorithms as in \cite{Mendoza2019a,DeAraujo2018a,Vuletic2016a}. 

Since optimal size and placement require a multi-objective optimization technique, and as discovered from the study, sequential process and priority-based objectives are the most applied form of multi-objective programming. The latter do require sensitivity analysis depending on the weight assessments. Recently, Pareto a posteriori optimal fronts have also been implemented for simultaneous independent solutions. Some of these algorithms are MOPSO, NSGA-II, and MOALO. Their implementation, especially with other techniques has yielded excellent results, taking a cue from \cite{RoyGhatak2018,Jani2018a,Li2018}.

Observations from the review show that some articles do not compare new algorithms results with existing ones. The authenticity of the data-set used may be questionable, because of the unknown source of the data-set. It is suggested that the application of any improved or newly developed meta-heuristic algorithm in the area of optimal sizing and placement, must be compared to previously applied algorithms with accompanying comments from results. Also, their evaluation must be carried out on a standard test system such as the IEEE bus and branch data. The trend of optimal sizing and placement of electrical units in a distribution network grid is shifting towards a full-blown smart grid, where clean energy will be prevalent. Hence, more studies will need to be carried out on optimal placement and sizing of ESSs, wind turbines, solar PVs, and PHEVs.

\begin{longtable}{>{\centering\arraybackslash}p{1.7cm}>{\centering\arraybackslash}p{1.2cm}>{\centering\arraybackslash}p{2cm}p{1.5cm}p{1.4cm}p{2cm}p{2cm}}
	\caption{Optimization with voltage stability indices\label{table2}}\\
	\hline\hline
	\multicolumn{7}{c}%
	{}\\
	\textbf{References} & \textbf{Grid scenario} & \textbf{MH algorithm} & \textbf{Flow model type} & \textbf{Placement type} & \textbf{Objective} & \textbf{Test system} \\
	\hline 
	\endfirsthead
	\caption[]{Optimization with voltage stability indices(continued)}\\
	\hline\hline
	\multicolumn{7}{l}%
	{}\\
	\textbf{References} & \textbf{Grid scenario} & \textbf{MH algorithm} & \textbf{Flow model type} & \textbf{Unit type} & \textbf{Objective} & \textbf{Test system} \\
	\hline
	\endhead
	\hline  \\
	\cite{AbulWafa2014a} & - & Fuzzy GA & OPF & Capacitors & power loss minimization & 33-node test feeder \\ 
	
	\cite{Moradi2016a} & MG & GA-IWD & OPF & DG & Power loss minimization & IEEE 33- and 69-bus system \\ 
	
	\cite{Poornazaryan2016a} & MG & ICA & OPF & DG & real power loss minimization, Voltage profile improvement & 34- and 69-bus  system \\ 
	
	\cite{El-Ela2018a} & - & WCA & OPF & DG and capacitors & real power loss minimization, Voltage profile improvement, cost reduction & IEEE 33- and 69-bus system. Egyptian grid \\ 
	
	\cite{Abdelaziz2016b} & DS & FPA & OPF & Capacitors & Power loss minimization and Voltage profile improvement & 15-, 69-, and 118-bus test systems \\
	
	\cite{Reddy2017a} & RE-based DS & WOA & B/F sweep PF & DG & Power loss minimization and Voltage profile improvement & IEEE 15-, 33-, 69- and 85-bus test systems \\
	
	\cite{Jani2018a} & WT-based grid & MOPSO-NSGA-II & PPF & BESS & Emission reduction, Voltage deviation improvement, cost reduction & IEEE 30-bus system \\ 
	
	\cite{Khaki2018} & WT-based grid & GA & PPF & WT-DG and BESS & Total power loss minization and cost reduction & IEEE 33-bus system \\ 
	
	\cite{Celli2018} & - & NSGA-II & PPF & BESS & Power loss minimization & - \\ 
	
	\cite{Zhang2018} & - & NSGA-II & PPF & DG & DG planning & - \\ 
	
	\cite{Mostafa2019} & RE-based grid & SOS & - & BESS & Power loss minimization, Voltage profile improvement & - \\
	
	\cite{Mahesh2016a} & DS & Advanced-PFNDMOPSO & B/F sweep PF & DG & Power loss minimization and Voltage stability improvement & IEEE 33-bus test systems \\	 
	
	\cite{Ali2016a} & DS & Improved HSA & OPF & Capacitors & Power loss minimization and total cost reduction & 15-, 69-, and 118-bus test systems \\	
	
	\cite{Lotfi2019} & - & IPSO-MSFLA &  &  &  &  \\ 
	
	\cite{RoyGhatak2018} & MG & E\_NSGA-II & PPF & PV, BESS, and STATCOM & Voltage profile improvement & 69-bus system \\ 
	\hline
\end{longtable} 

\begin{longtable}{>{\centering\arraybackslash}p{1.7cm}>{\centering\arraybackslash}p{1.2cm}>{\centering\arraybackslash}p{2cm}p{1.5cm}p{1.4cm}p{2cm}p{2cm}}
	\caption[An optional table caption ...]{Optimization with Loss Sensitivity Factors\label{table3}}\\
	\hline\hline
	\multicolumn{7}{c}%
	{}\\
	\textbf{References} & \textbf{Grid scenario} & \textbf{MH algorithm} & \textbf{Flow model type} & \textbf{Placement type} & \textbf{Objective} & \textbf{Test system} \\
	\hline 
	\endfirsthead
	\caption[]{Optimization with Loss Sensitivity Factors(continued)}\\
	\hline\hline
	\multicolumn{7}{l}%
	{}\\
	\textbf{References} & \textbf{Grid scenario} & \textbf{MH algorithm} & \textbf{Flow model type} & \textbf{Placement type} & \textbf{Objective} & \textbf{Test system} \\
	\hline
	\endhead
	\hline 
	\cite{Elsheikh2014a} & - & Discrete PSO & OPF & Capacitors & Power loss minimization and voltage profile improvement & 10-, 15-, and 34-bus system \\ 
	
	\cite{Diab2017a} & - & Hybrid PSO & B/F sweep & Capacitors & Power loss minimization and voltage profile improvement & IEEE 15- and 34-bus system. Moscow grid \\ 
	
	\cite{Mendoza2019a} & - & PSO & B/F sweep & Capacitors & Power loss minimization and voltage profile improvement & 34 and 85-bus system \\ 
	
	\cite{Sharma2019a} & - & ABC-PSO & Fuzzy load flow & Capacitors & Power loss minimization and voltage profile improvement & 34-node test feeder \\ 
	
	\cite{MohamedShuaib2015a} & - & GSA & NR-Fast decoupling LF & Capacitors & Power loss minimization  & 33-,69-,85-, and 141-bus system \\ 
	
	\cite{Abdelaziz2016c} & - & FPA & B/F sweep & Capacitors & Power loss minimization  & 10-, 33-, and 69-bus system \\ 
	
	\cite{Upper2018a} & - & MFA & B/F sweep & Capacitors & Energy loss minimization  & 69-bus system \\ 
	
	\cite{Ali2018a} & RE-based grid & ALO & - & DG & Power loss minimization and savings increase & 69-bus system \\ 
	
	\cite{Saha2016a} & MG & Chaos SOS & - & DG & Power loss minimization and voltage profile improvement & 33-,69-, and 118-bus system \\ 
	
	\cite{Kishore2018a} & RDNs & IBFOA & Fuzzy laod flow & Capacitors & Power loss minimization and voltage profile improvement & 33-,69-, and 141-bus system \\ 
	
	\cite{Li2018} & Active RDNs & MOALO & PPF & BESS and DG & Power loss minimization and voltage profile improvement &  \\ 
	
	\cite{Babacan2017} & Active RDNs & GA & - & BESS & Power loss minimization and voltage fluctuation reduction & IEEE 8500-node \\ 
	\hline 
\end{longtable} 

\section{Conclusion}\label{section5}
Distributed generation units, capacitors, and D-STATCOM have played an important role for distribution systems in power system. These developments have brought a new paradigm to the power grid. For instance, active power flow is present compared to the traditional passive power flow, where a power generation only comes from one feed-source. Also, there have been improvements in voltage stability and minimized power loss, by optimally allocating and sizing DGs and other electrical units. The study reveals that power loss minimization, voltage profile improvement, and cost minimization are the most common objectives while finding optimal location and size. 

This study has reviewed the application of meta-heuristic algorithms for solving the optimal placement and sizing problem, and also its dynamic implementation to solve objective functions. These algorithms have evolved into new and improved ones, thereby making room for new improvements in smart grids. Since the optimal location and size problem is recently based on improving more than one objective, researchers are faced with an additional decision making, which is to choose a convenient but correct method to handle objective functions. It is noteworthy that the handling of objective functions correctly can be equivalent to the authenticity of results. 

Some common methods of handling multiple objective functions in the optimal placement and sizing problem are sequential handling and priority-based handling. The former is the simplest form but comes with a limitation. To correctly utilize this method, objective functions must be closely related. In most cases, the solution of an objective function can be a variable in another function.  Priority-based handling is the most common method. Researchers have to assign weights (prior to the optimization process) to objective functions based on (i) their solid understanding of power systems or (ii) previous studies. This process seem subjective since there will always be human intervention even after developing objective functions. Another method is the Pareto-based handling. This is a recent, independent, and non-deterministic approach to handling multiple objective functions. The set of objectives are optimized at the same time, to find non-dominated solutions from each objective. Future research in this direction will domain will be interesting.

Since optimization algorithms may require a high complexity for solving the optimal placement problem, model simplification method has been widely used. This method uses a two-step framework which firstly derive a simple model (or a bus network) and apply an algorithm to solve a problem based on the simplified network. Techniques such as LSF, VSI, and PLI have been used to simplify bus networks. These techniques are based on different traditional power system calculations and may come at a cost of information loss, hence it affects the overall performance of the algorithm. There are two prospective solutions to improve the overall performance. First is the development of better techniques to improve optimal location. Second is the improvement of the efficiency of metaheuristic algorithms to handle a whole bus network. Overall, balancing the complexity (of bus networks) and the efficiency (of meta-heuristic algorithms) is very important. 

However, suggested future works may be required in the
\begin{itemize}
    \item development of better objective functions to explain the features of a distribution networks
    \item development of better techniques for selecting candidate buses from a bus network
    \item development of efficient algorithms to conveniently solve the optimal placement and sizing problem
	\item development of multi-objective based meta-heuristic algorithms for non-dominated solutions for optimal placement and sizing problem.
\end{itemize}


\bibliographystyle{unsrt}      
\bibliography{references4}   

\end{document}